\documentclass[12pt]{iopart}
\usepackage{iopams,harvard,bm,graphicx}  
\begin{document}

\title[]{Periodic orbits in Large Eddy Simulation of Box Turbulence}

\author{Lennaert van Veen$^1$\footnote{Corresponding 
author: lennaert.vanveen@uoit.ca}, Alberto Vela Mart\'in$^2$, Genta Kawahara$^3$ and Tatsuya Yasuda$^4$}

\address{$^1$Faculty of Science, University of Ontario Institute of Technology, Oshawa, Ontario, Canada}
\address{$^2$School of Aeronautics, Universidad Polit\'ecnica de Madrid, 28040 Madrid, Spain}
\address{$^3$Faculty of Engineering Science, Osaka University, Toyonaka, Osaka, Japan}
\address{$^4$Faculty of Engineering, Department of Aeronautics, Imperial College, London, United Kingdom}


\begin{abstract}
We describe and compare two time-periodic flows embedded in Large Eddy Simulation (LES) of turbulence in a three-dimensional, 
periodic domain subject to constant external forcing. One of these flows models the regeneration of large-scale structures that was observed 
in this geometry by Yasuda et al. ({\sl Fluid Dyn. Res.} {\bf 46}, 061413, 2014), who used a smaller LES filter length and thus obtained 
a greater separation of scales of coherent motion. We speculate on the feasibility of modelling the regenerative dynamics with time-periodic
solutions in such a flow, which may require novel techniques to deal with the extreme ill-conditioning of the associated boundary value problems.  
\end{abstract}

\vspace{2pc}

\maketitle

\section{Introduction}

One of the most important open questions in the study of Homogeneous Isotropic Turbulence (HIT) is that of the dynamical processes that govern the energy cascade process
in the inertial subrange. The theory of Kolmogorov\citeyear{K}, based on similarity hypotheses and dimensional analysis, offers only a statistical description of the transfer of 
energy from the largest scales, on which external forces act, down to the smallest scales, on which it is dissipated. Various mechanisms have been proposed by which coherent
structures in physical space could interact to give rise, on aggregate, to a scale-invariant energy spectrum (see, e.g., \citeasnoun{G} and references therein).
Establishing the relevance of such mechanisms is quite challenging, however, due to the spatio-temporal complexity of turbulent flows with a developed inertial subrange
as well as their extremely sensitive dependence on initial conditions.

This difficulty prompted \citeasnoun{YGK} to systematically investigate inertial range dynamics in a highly constrained flow, keeping the complexity down to the
bare minimum. Firstly, they considered periodic boundary conditions only, ruling out boundary layers and the associated multiscale phenomena. 
Secondly, they imposed an external body force constant in time 
and with a simple spatial structure, rather than forcing the flow in a time-dependent manner designed to maintain statistical isotropy and homogeneity on all scales. 
Thirdly, they employed Large Eddy Simulation (LES) to simulate intertial range dynamics on relatively small grids, thereby greatly reducing the computational demands.

Their main result was that, even in the presence of a developed inertial range, the dynamics was quasi-cyclic. On grids ranging from $16^3$ to $256^3$, with the LES filter length
fixed to the grid spacing, large-amplitude fluctuations in the kinetic energy and the rate of energy dissipation (or rather transfer to sub grid scales) were observed. These quasi-cyclic
fluctuations hint at the existence of a regeneration cycle of large-scale coherent structures. Such a regeneration cycle could be explained in terms of the interaction of
vortices on different spatial scales, and thus in terms of energy transfer.

It would be desirable to compute these large-scale fluctuations as time-periodic solutions -- a proposition left for future research by Yasuda and coworkers. 
Periodic solutions that capture the regeneration cycle would enable us to study the vortex interaction mechanism in detail and explore its dependence on the range of active spatial
scales. The current paper reports on  a first attempt to attain this goal. Using the LES filter length as a control parameter on a fixed grid of $64^3$ points, we identified two
unstable periodic solutions with a different character. One has a short period, a small amplitude and relatively high energy while the other has a long period and exhibits a fluctuation
of spatial mean quantities comparable to those of turbulence motion. The short periodic orbit appears to be situated in between the turbulent and the laminar flows in phase space, somewhat 
similar to the ``quiescent'' periodic orbit in planar Couette flow described by \citeasnoun{KawKida01}. We present these solutions at a value of the LES filter length 
only about sixteen times smaller than the integral length scale and far greater than the grid spacing. At this scale separation, only the beginning of a power law can be observed in
the energy spectrum. In the concluding section we discuss the obstacles that can be expected when computing invariant solutions in LES with a developed Kolmogorov spectrum.

\section{LES with Taylor-Green forcing}

The momentum balance and continuity equations for LES with the Smagorinsky model are
\begin{eqnarray}\label{LES}
\partial_t\bm{u}+\bm{u}\!\cdot\!\nabla \bm{u} +\nabla \left(\frac{p}{\rho}+\frac{1}{3}\Pi\right) -2\nabla(\nu_T\bm{S})=\gamma \bm{f} \\
\nabla\!\cdot\! \bm{u}=0
\end{eqnarray}
where $\bm{u}$, $p$ and $\bm{S}$ are the filtered velocity, pressure and rate-of-strain tensor, respectively, and $\Pi$ contains the normal sub grid stress. The density, $\rho$, and the
amplitude of the external force, $\gamma$, are constant and the molecular viscosity has been set to zero. Using the closure
proposed by \citeasnoun{S}, the eddy viscosity is
given by 
\begin{equation}\label{Smodel}
\nu_T=(C_{\rm S} \Delta)^2 \sqrt{2 S_{ij} S_{ij}}
\end{equation}
where $C_{\rm S}$ is the nondimensional Smagorinsky parameter and $\Delta$ is the grid spacing. Summation over repeated indices is implied.
We fix the external force on a periodic domain of dimensions $L\times L\times L$ to $\bm{f}=(-\sin(k_{\rm f}x)\cos(k_{\rm f}y),\cos(k_{\rm f}x)\sin(k_{\rm f}y),0)^t$ with $k_{\rm f}=2\pi/L$. This body force induces a flow with four counter-rotating vortex columns,
which is one of a family of flows studied by \citeasnoun{TG}.

In the following, we nondimensionalize the equations according to
\begin{eqnarray}\label{nondim}
\bm{x}'&=k_{\rm f}\,\bm{x}      \\
t'&=\sqrt{L\gamma k^2_{\rm f}}\,t \\
\bm{u}'&=\sqrt{\frac{1}{L\gamma}}\,\bm{u} \\
p'&=\frac{1}{\rho L\gamma}\,p\\
\Pi'&=\frac{1}{L\gamma}\,\Pi
\end{eqnarray}
and drop the primes for non dimensional quantities. The resulting momentum balance equation is
\begin{equation}\label{nondimLES}
\bm{u}_t+\bm{u}\!\cdot\! \nabla \bm{u}+\nabla \left(p+\frac{1}{3}\Pi\right)-\frac{2}{R^{3/2}}\nabla \left(\sqrt{2S_{ij}S_{ij}}\bm{S}\right)=\frac{1}{\alpha} \bm{f} 
\end{equation}
with non dimensional parameters
\begin{eqnarray}\label{Reyal}
R&=\left(\frac{1}{C_{\rm S}\Delta k_{\rm f}}\right)^{4/3} \\
\alpha&=L k_{\rm f}=2\pi
\end{eqnarray}
where we followed \citeasnoun{M} in defining the LES Reynolds number $R$.
The kinetic energy is given by
\begin{equation}\label{en_ens}
K=\frac{1}{\alpha^3}\int \frac{1}{2}\|\bm{u}\|^2\,\mbox{d}\bm{x}=\int\limits_{0}^{\infty} E(k)\,\mbox{d}k
\end{equation}
where $E(k)$ is the energy spectrum. The integral length scale is computed as
\begin{equation}
\ell = \left<\!\frac{3\pi}{4K} \int\limits_{0}^{\infty} \frac{E(k)}{k}\,\mbox{d}k\!\right>
\label{ILS}
\end{equation}
and the integral time scale as $T=\ell/\sqrt{2<\!K\!>/3}$, where $<.>$ denotes the space-time or ensemble average.
The rate of energy transfer to sub grid scales is given by
\begin{equation}\label{EDR}
\epsilon=\frac{2}{\alpha^3 R^{3/2}}\int \sqrt{2S_{ij}S_{ij}} S_{k\ell}S_{k\ell}\,\mbox{d}\bm{x}
\end{equation}
and the rate of energy input by
\begin{equation}\label{EIR}
e=\frac{1}{\alpha^4} \int \mathbf{u}\!\cdot\!\mathbf{f}\,\mbox{d}\bm{x}
\end{equation}
 
\begin{figure}
\begin{center}
\includegraphics[width=0.6\textwidth]{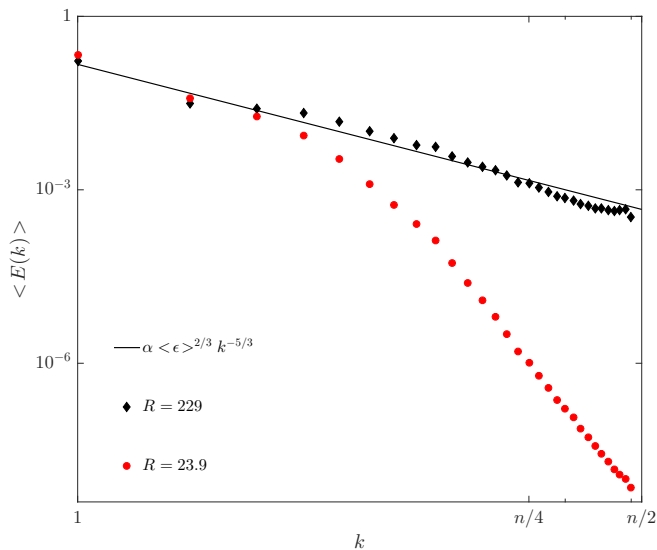}
\end{center}
\caption{Three-dimensional energy spectrum for the Smagorinsky LES flow with four-vortex forcing. The black line denotes Kolmogorov's theoretical prediction for the 
inertial range spectrum, while the symbols denote the spectra for $R=229$ ($C_{\rm S}=0.173$) in black diamonds and $R=23.9$ 
($C_{\rm S}=0.943$) in red dots.}
\label{spectra}
\end{figure} 

We simulate the flow and the tangent linear model with a Fourier pseudo-spectral code at a resolution of $64^3$ grid points. 
De-aliasing of the quadratic nonlinearity is accomplished by the phase-shift method of \citeasnoun{PO}. 
The largest resolved wave number in this method is $\lfloor 0.94 \times n/2 \rfloor=30$, where $n/2=32$ is the Nyquist wave number.
Time is discretized using a fourth order Runge-Kutta-Gill scheme with a step size of 0.06 in nondimensional units.
The actual degrees of freedom in the resulting discretized system are the $N=230240$ Fourier coefficients of two components of vorticity. 
A detailed description of this $N$-dimensional dynamical system, including all of its symmetries, can be found in \citeasnoun{VKY}.
Here, we will only state the fact that the system is equivariant under translations along the vertical direction, i.e. the $z$-axis. The spatial mean flux in this direction
is fixed to zero in the simulations, and the invariant solutions presented below are periodic relative to a shift along the vertical direction.

In Fig. \ref{spectra} the energy spectrum on turbulent flow is shown for two different LES Reynolds numbers, along
with the prediction from Kolmogorov's theory \cite{K}. For the latter we have assumed that the constant of proportionality $\alpha=1.5$ and $<\!\epsilon\!>$ was
fixed to the spatio-temporal average of the transfer rate at $R=229$. The latter LES Reynolds number corresponds to $C_{\rm S}=0.173$, the value Lilly estimated to 
produce a $-5/3$ power law spectrum for all wave numbers \cite{Lilly}. 
Indeed, the measured spectrum at this control parameter approximately follows this power law. The other spectrum shown corresponds to the
LES Reynolds numbers at which we computed periodic solutions. Only the beginning of a power law spectrum is visible here as the energy input scale
and filter scale $C_{\rm S}\Delta$ are separated only by a factor of about sixteen.

\section{Unstable periodic motion}

We computed two invariant solutions of very different periods using conventional methods. An initial guess for each solution was filtered from a turbulent time series
by recording spontaneous near-recurrences. Subsequently, we used the Newton-Krylov-hook algorithm as proposed by \citeasnoun{vis} to iteratively refine the solutions.
Nearly 400 iterations were performed on each solution with a GPU-accelerated code over the course of several months. For the solution of short period, the resulting
residual is of order $10^{-6}$ as measured in the energy norm, normalized by the space and time averaged energy.
For the solution of long period, the resulting
residual is of the order $10^{-4}$ in the same norm.
\begin{figure}
\begin{center}
\includegraphics[width=0.6\textwidth]{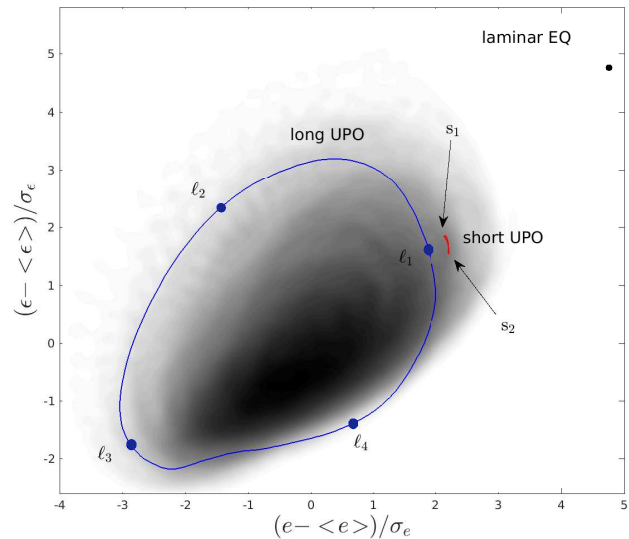}
\end{center}
\caption{Probability density function of turbulence projected onto $e$ and $\epsilon$ with superimposed the laminar equilibrium and two unstable periodic solutions. All solutions were computed at $R=23.9$. The labels ${\rm s}_1$, ${\rm s}_2$ and $\ell_1\ldots \ell_6$ correspond to the physical space portraits shown in figures \ref{short_UPO} and \ref{long_UPO}.}
\label{PDF_UPO}
\end{figure} 

\begin{figure}
\begin{center}
\includegraphics[width=0.47\textwidth]{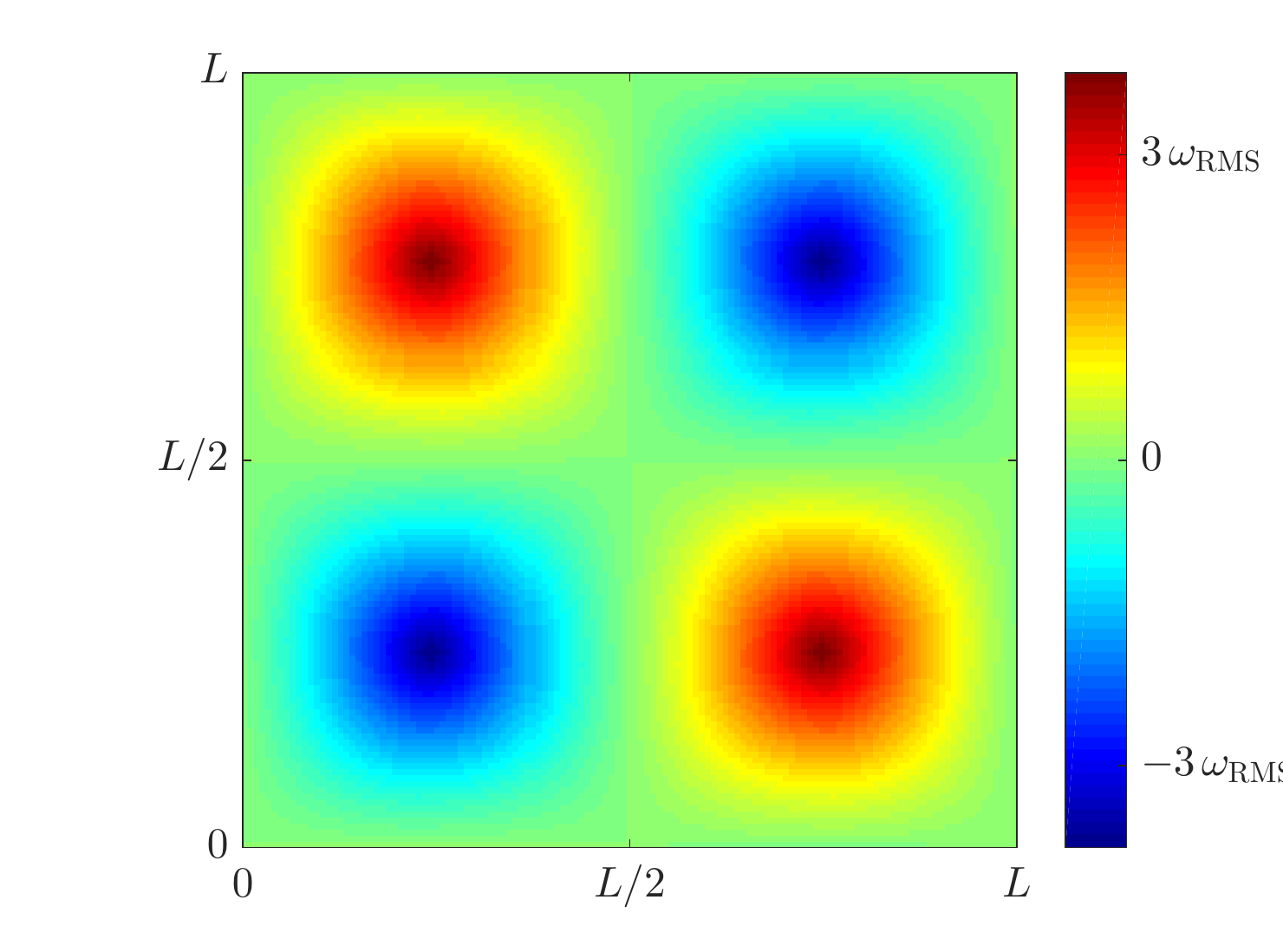} 
\end{center}
\caption{Physical space portrait of the laminar equilibrium. Shown is the vertical vorticity in a plane of constant height $z$. The root-mean-square value of the vertical
vorticity is indicated on the scale.}
\label{laminar}
\end{figure}

\begin{figure}
\begin{center}
\includegraphics[width=0.47\textwidth]{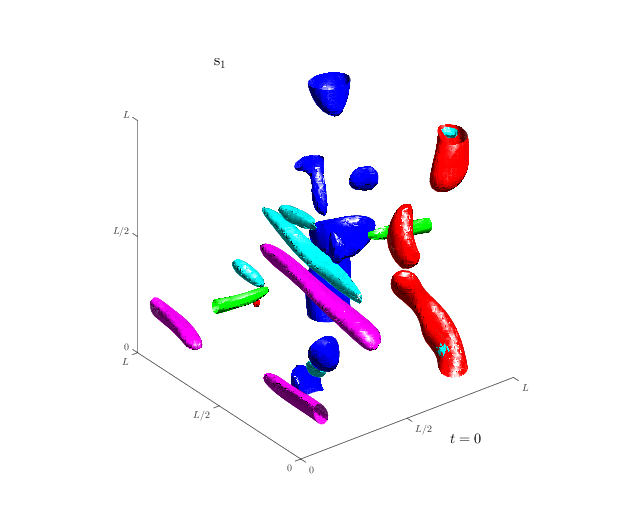} \includegraphics[width=0.47\textwidth]{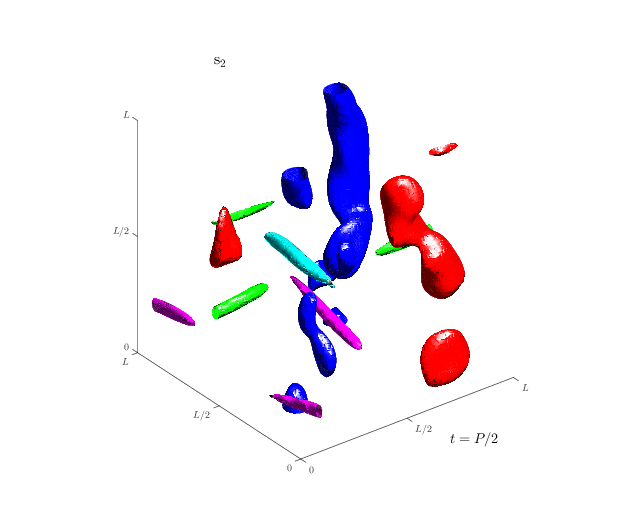}
\end{center}
\caption{Physical space portraits of the short periodic solution. The labels corresponds to those in figure \ref{PDF_UPO}. Shown are the isosurfaces of $z$-vorticity (red and blue for positive and negative), $x$-vorticity (green and yellow for positive and negative) and $y$-vorticity (cyan and magenta  for positive and negative) at $75\%$ of the maximal and minimal values.}
\label{short_UPO}
\end{figure}

Figure \ref{PDF_UPO} shows the resulting solutions superimposed on the probability density function of turbulent motion at $R=23.9$, projected onto the rate of energy input and transfer to subgrid scales. 
Also included is the laminar equilibrium. The latter is translation invariant in the vertical direction, so we can portray
it using the vertical vorticity as shown in figure \ref{laminar}. The instabilities of this equilibrium were examined in detail by \citeasnoun{VKY}. In that work, a long period orbit is computed at $R=14$, and at that LES Reynolds number the laminar equilibrium is approached occasionally in the ambient turbulent motion. At the current value of $R$, an unstable relative periodic solution exists in between the laminar and the turbulent 
flows in phase space. It has a period of $2.4T$ and in each period shifts by a distance of about $L/12$ in the vertical direction. Two snap shots are shown in figure \ref{short_UPO}. 
The vortex columns that dominate
the laminar solution have been deformed and several pairs of counter-rotating smaller-scale vortices have been created, most notably in the centre of the domain. The latter vortices vary in 
intensity over the course of one period but remain almost stationary. The time series of the kinetic energy of this solution is shown in figure \ref{time_series}. The turbulent motion occasionally
comes close to the periodic solution, that appears to play a role similar to the ``quiescent'' periodic orbit identified by \citeasnoun{KawKida01} in transitional Couette flow. However, the
current solution has multiple unstable multipliers -- at least four -- and thus does not separate qualitatively different regions of phase space in the way edge states do in sub 
critical transition scenarios.

\begin{figure}[p]
\begin{center}
\includegraphics[width=0.47\textwidth]{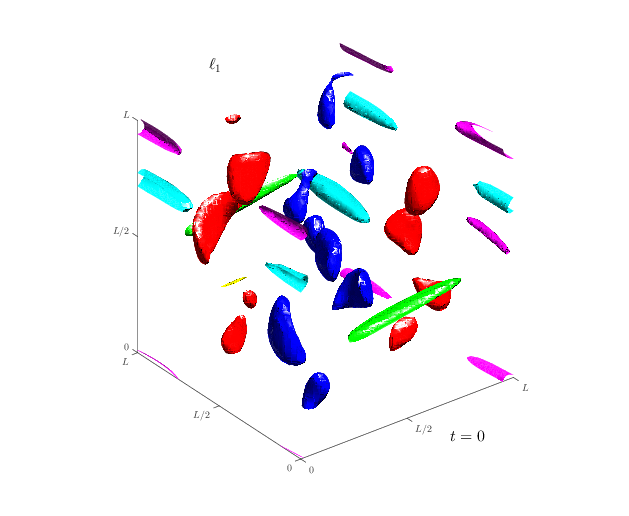} \includegraphics[width=0.47\textwidth]{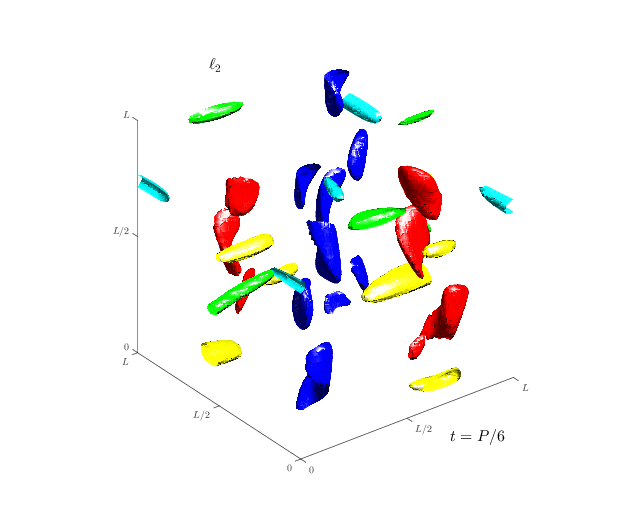}\\
\includegraphics[width=0.47\textwidth]{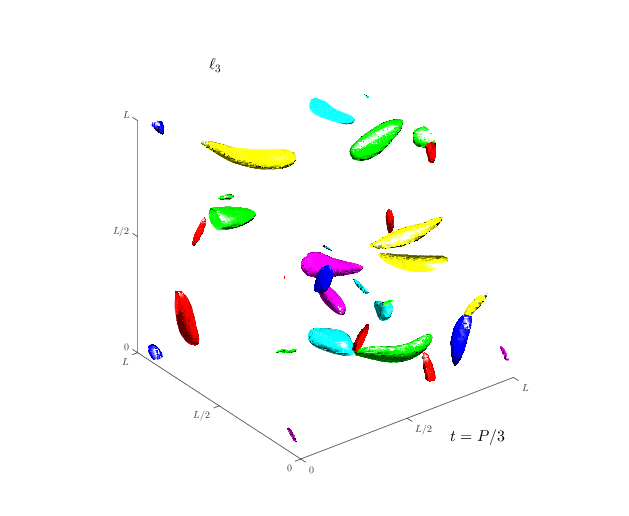} \includegraphics[width=0.47\textwidth]{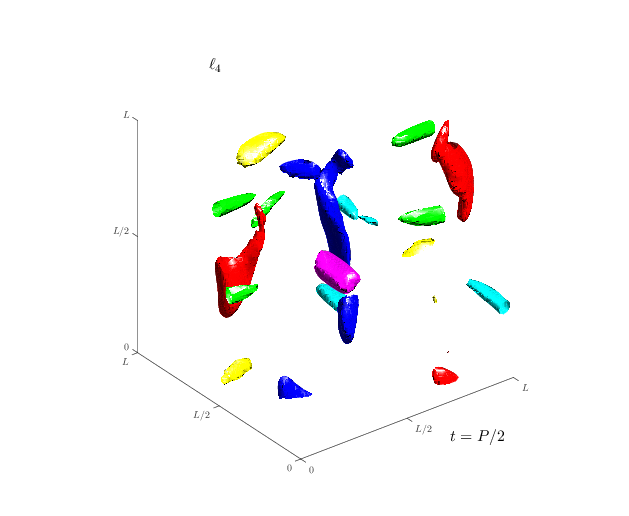}\\
\includegraphics[width=0.47\textwidth]{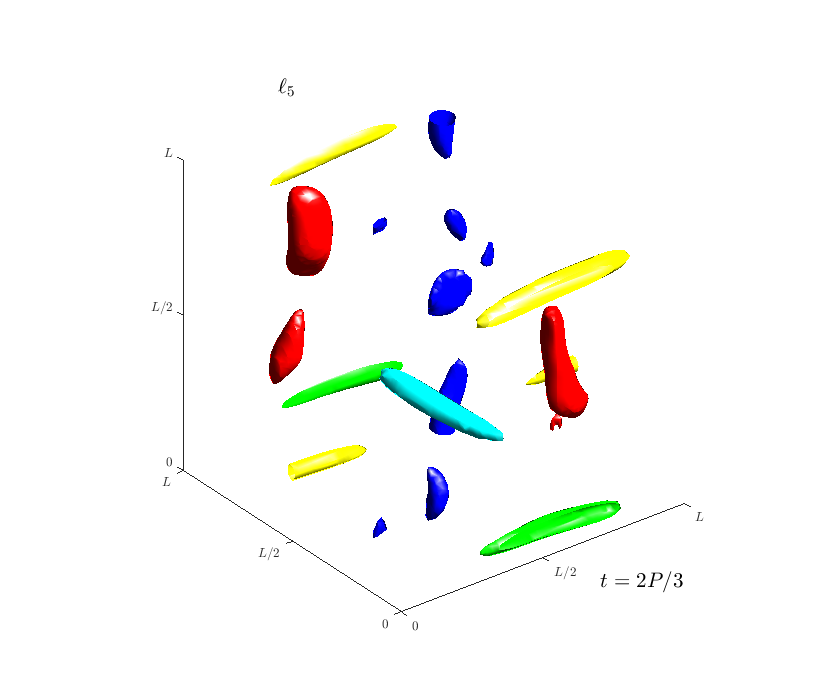} \includegraphics[width=0.47\textwidth]{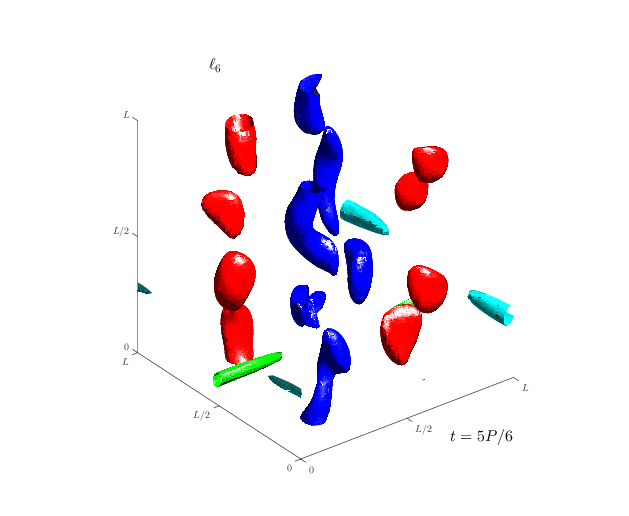}
\end{center}
\caption{Physical space portraits of the long periodic solution. The labels corresponds to those in figure \ref{PDF_UPO}. The colour coding is the same as in figure \ref{short_UPO}.}
\label{long_UPO}
\end{figure}

\begin{figure}
\begin{center}
\includegraphics[width=0.6\textwidth]{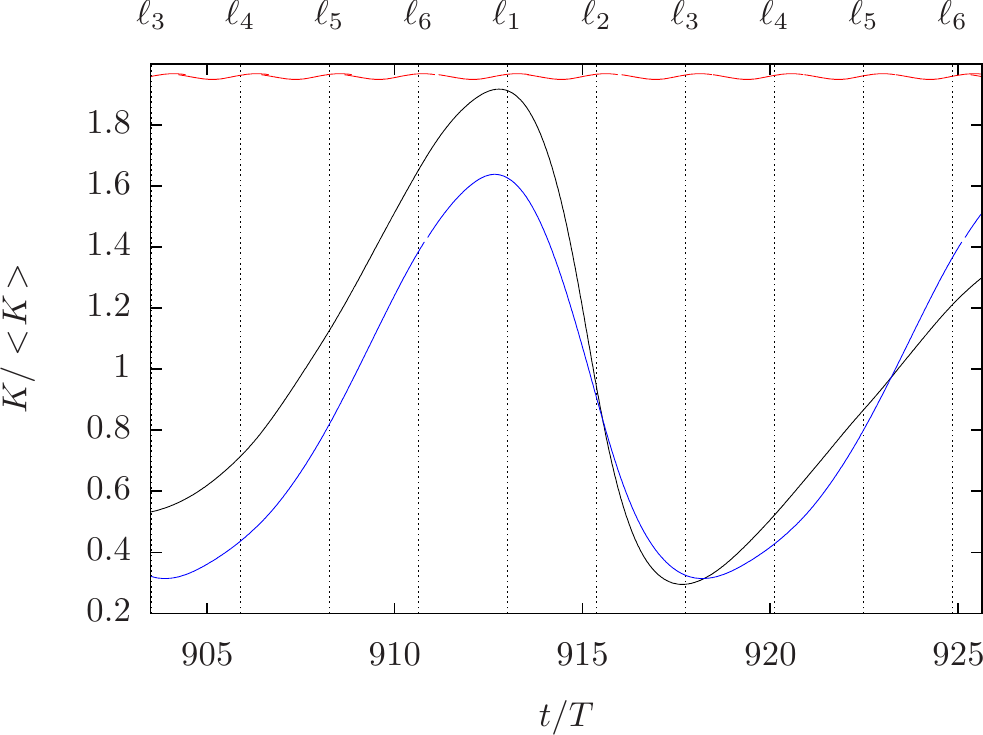}
\end{center}
\caption{Time series of the short periodic orbit (red), the long periodic orbit (blue) and a segment of freely evolving turbulence (black). Shown is the kinetic energy, normalized by its spatio-temporal mean, versus time, normalized by the integral time scale. The labels at the top correspond to the snap shots in figure \ref{long_UPO}. This figure can be compared to figure 5 from \citeasnoun{YGK}.}
\label{time_series}
\end{figure}

The solution of longer period is shown in blue in figure \ref{PDF_UPO}. Its period of $14.2T$ is close to that of the spontaneous, large-amplitude oscillations observed
by \citeasnoun{YGK} at higher LES Reynolds numbers. Indeed, in the projection onto energy input and transfer rate it appears to shadow such oscillations. The variation of the energy
along the orbit similarly compares well to that in turbulent motion, as shown in figure \ref{time_series}. Six snap shots, taken at equal intervals along the solution, are shown
in figure \ref{long_UPO}. At time $t=0$, the energy input rate and the energy are large. The large-scale vortex columns excited by the forcing are partly visible and, in addition, pairs of counter-rotating 
vortices are generated in their straining field. In the second snap shot, the large-scale vortices have weakened, corresponding to a decreasing energy input rate. The activity on
smaller scales has increased and the rate of energy transfer to sub grid scales is high. After a third of the period, the energy input is minimal and no large-scale structure is 
present. Subsequently, the small-scale structures lose energy while, at the same time, the large-scale vortex columns start to regain strength due to the external forcing.
This cyclic behaviour is similar to the behaviour described by \citeasnoun{YGK}, even though the flows they studied have a much greater separation of length scales. 
This gives some hope that it may be possible to capture the regenerative dynamics by time-periodic solutions even at much higher LES Reynolds numbers.

\section{Outlook}

We have demonstrated that, at least at a relatively low LES Reynolds number, time-periodic solutions can be computed that capture the regeneration of large-scale structures
and the associated oscillations in the energy and the rate of its input and transfer to sub grid scales. The ratio of the integral length scale to the LES
filter length, $\ell/(C_{\rm S}\Delta)$, is only about 16 in this computation. Nevertheless, the number of Newton-hook iterations performed on each of the solutions was in the hundreds.
Each of these iterations required around 1000 Krylov subspace iterations and the whole computation took several months on GPU facilities. 
Since the ultimate goal is to repeat this computation at an LES Reynolds number about ten times larger, it is worthwhile to speculate about the feasibility of this approach. 

While LES drastically reduces the number of DOF in turbulent simulations, it does not yield boundary value problems that are conducive to solving by the conventional Newton-Krylov-hook algorithm.
As pointed out by \citeasnoun{sanch}, the efficiency of this algorithm relies on the clustering of the eigenvalues of the monodromy matrix which, in turn, is caused by the dominance of the viscous damping on small length scales.
The DOF resolved in LES with a Kolmogorov spectrum are not clustered and, as a consequence, many Krylov iterations are necessary to find the Newton update steps. If we make the simple assumption that the number of necessary Krylov iterations is proportional to the number of active DOF, we expect it to grow as $2/(C_{\rm S} \Delta k_{\rm f})^3$. That would mean that increasing the LES Reynolds number to 100 would require an increase by a factor of 60 in the number of Krylov iterations. This will likely lead to an accumulation of round-off error, accrued in time-stepping and orthogonalization of the Krylov basis, that ultimately limits the convergence. Moreover, the computation time would be in the order of years.

In summary, the computation of invariant solutions with a small but significant Kolmogorov spectrum may be feasible, and is the goal of ongoing research. In order to study LES with a Kolmogorov spectrum spanning more than one decade, it will probably be necessary to develop new techniques for preconditioning and solving the relevant boundary value problems.

\ack
LvV was supported by an Individual Discovery Grant of NSERC. We gratefully acknowledge funding by the Coturb program of the European Research Council.

\section*{References}
\bibliography{POLESBT}
\bibliographystyle{jphysicsB}

\end{document}